\documentclass[
 preprint,
superscriptaddress,
 amsmath,amssymb,
 aps,
prl,
]{revtex4-2}

\usepackage{physics}
\usepackage{booktabs}
\usepackage{braket}
\usepackage{qcircuit}
\usepackage{graphicx}
\usepackage{natbib}
\usepackage{dcolumn}
\usepackage{bm}
\usepackage{makecell}
\usepackage{tensor}
\usepackage{xcolor}
\usepackage{threeparttable}
\usepackage{chemformula}
\counterwithin{subsection}{section}
 
\makeatletter
\def\switch@array{}
\makeatother

\begin{document}

\title{AdsMind: A Physics-Grounded Multi-Agent System for Self-Correcting Discovery of Adsorption Configurations on Heterogeneous Catalyst Surfaces}

\author{Zongmin Zhang}
\affiliation{Department of Computer Science and Engineering, Hong Kong University of Science and Technology, Kowloon, Hong Kong 999077, China}
\author{Yuyang Lou}
\affiliation{Department of Chemistry, Hong Kong University of Science and Technology, Kowloon, Hong Kong 999077, China}
\author{Bowen Zhang}
\affiliation{Department of Chemistry, Hong Kong University of Science and Technology, Kowloon, Hong Kong 999077, China}
\author{Junwu Chen}
\affiliation{Laboratory of Artificial Chemical Intelligence (LIAC), 
  Institute of Chemical Sciences and Engineering, 
  Ecole Polytechnique F\'{e}d\'{e}rale de Lausanne (EPFL), 
  Lausanne 1015, Switzerland}
\author{Ryo Kuroki}
\affiliation{Laboratory of Artificial Chemical Intelligence (LIAC), 
  Institute of Chemical Sciences and Engineering, 
  Ecole Polytechnique F\'{e}d\'{e}rale de Lausanne (EPFL), 
  Lausanne 1015, Switzerland}
\affiliation{Platform Laboratory for Science \& Technology, Asahi Kasei Corporation, Tokyo, Japan}
\author{Xuan Vu Nguyen}
\affiliation{Laboratory of Artificial Chemical Intelligence (LIAC), 
  Institute of Chemical Sciences and Engineering, 
  Ecole Polytechnique F\'{e}d\'{e}rale de Lausanne (EPFL), 
  Lausanne 1015, Switzerland}
\author{Edvin Fako}
\email{edvin.fako@epfl.ch}
\affiliation{Laboratory of Artificial Chemical Intelligence (LIAC), 
  Institute of Chemical Sciences and Engineering, 
  Ecole Polytechnique F\'{e}d\'{e}rale de Lausanne (EPFL), 
  Lausanne 1015, Switzerland}
\author{Lixue Cheng}
\email{lixuecheng@ust.hk}
\affiliation{Department of Chemistry, Hong Kong University of Science and Technology, Kowloon, Hong Kong 999077, China}
\affiliation{IAS Center for AI for Scientific Discoveries, Hong Kong University of Science and Technology, Kowloon, Hong Kong 999077, China}
\author{Philippe Schwaller}
\email{philippe.schwaller@epfl.ch}
\affiliation{Laboratory of Artificial Chemical Intelligence (LIAC), 
  Institute of Chemical Sciences and Engineering, 
  Ecole Polytechnique F\'{e}d\'{e}rale de Lausanne (EPFL), 
  Lausanne 1015, Switzerland}
\date{\today}

\begin{abstract}

Identifying the lowest-energy surface--adsorbate configuration is critical for modeling heterogeneous catalysis, yet exhaustive exploration with ab initio calculations is computationally prohibitive. Machine-learning force fields (MLFFs) accelerate structural relaxation but leave the search over the vast configurational space a major bottleneck, and open-loop large language model (LLM) agents lack a physics-grounded feedback mechanism to correct erroneous initial guesses.
We propose \textbf{AdsMind} (\textbf{Ads}orption configuration discovery with \textbf{M}achine \textbf{in}telligence and relaxation fee\textbf{d}back), a closed-loop multi-agent framework that enables autonomous error correction through MLFF relaxation feedback. Across four LLM backends, AdsMind achieves consistently high search reliability, with success rates of 100\% and 98.8\% on the benchmarks AA20 and OCD-GMAE62. Relative to its single-pass (1-Shot) ablation it reduces cross-backend energy dispersion, and it uses only 4.11 and 4.67 MLFF relaxations per case, respectively---an approximately 14-fold reduction over heuristic enumeration baselines.
Density functional theory (DFT) validation using VASP/PBE on six representative AA20 systems shows that the reported open-loop Adsorb-Agent outputs exhibit qualitative adsorption-energy sign errors for molecular adsorbates, whereas AdsMind preserves the correct sign in all tested cases with closer quantitative agreement. AdsMind thus delivers reliability, self-reflection, and interpretability simultaneously, supporting more DFT-informed autonomous chemistry workflows.

\end{abstract}

\maketitle 

\section{Introduction}

Powering most chemical manufacturing, heterogeneous catalysis is central to the clean-energy transition, enabling green hydrogen production, CO$_2$ valorization, and sustainable ammonia synthesis~\cite{norskov2011density}. Density functional theory (DFT) has accelerated computational catalyst discovery relative to experimental trial-and-error, yet translating predictions into practical catalysts remains difficult because of the accuracy limits of DFT and the combinatorial complexity of catalyst surfaces~\cite{norskov2011density,ulissi2017toward}. In every catalytic cycle, adsorption of molecular species onto the surface is an elementary step whose energetics govern activity and selectivity and underpin the microkinetic models and volcano plots used to select catalysts~\cite{norskov2011density,andersen2021adsorption}. Accurately identifying the most stable surface--adsorbate configuration is therefore central to computational catalyst discovery, from descriptor-based screening~\cite{andersen2021adsorption} to high-throughput DFT workflows~\cite{greeley2017design,ulissi2017toward,yeo2021high,rosen2019identifying}. This identification is itself a persistent bottleneck: the configuration space grows combinatorially with site type (ontop, bridge, hollow, and alloy variants), adsorbate orientation, and conformational flexibility, easily exceeding several hundred candidates for a single polyatomic adsorbate on an intermetallic surface and reaching thousands once co-adsorption or solvation is considered. Existing strategies each carry trade-offs: graph-theoretic enumeration of multidentate configurations~\cite{deshpande2020graph} and chemical-environment-based machine learning~\cite{ghanekar2022adsorbate} narrow the search; exhaustive DFT over canonical site families is systematic but prohibitively expensive, as each relaxation costs hundreds to thousands of CPU-hours and a single surface can demand $10^2$--$10^3$ evaluation cycles; and stochastic search methods---genetic algorithms~\cite{vilhelmsen2014genetic}, basin hopping~\cite{wales1997global}, \emph{ab initio} random structure searching~\cite{pickard2011ab}, and kinetic Monte Carlo~\cite{stamatakis2011first}---reduce evaluations relative to exhaustive enumeration but still demand substantial expert oversight and scale poorly across composition space~\cite{ulissi2017toward}.

Machine-learning force fields (MLFFs) have sharply reduced per-configuration cost: architectures such as MACE/MACE-MP~\cite{batatia2022mace,batatia2023foundation}, CHGNet~\cite{deng2023chgnet}, and EquiformerV2~\cite{liao2024equiformerv2} reach near-DFT accuracy at orders-of-magnitude speedup~\cite{lan2023adsorbml}, and large benchmarks such as OC20~\cite{chanussot2021open} and OC22~\cite{tran2023oc22} have standardized adsorption-energy evaluation. MLFFs, however, address only the \textit{fast energy and force evaluation} problem, not the upstream \textit{search-strategy} problem of deciding which configurations to evaluate. Search strategies are either \textit{open-loop}, with the policy fixed before evaluation, or \textit{closed-loop}, with each result fed back to adapt subsequent proposals. Most deep-learning approaches are open-loop: graph neural networks predict adsorption energies in a single pass~\cite{reiser2022gnn,adsmt,chowdhury2018prediction}, and generative crystal-structure models propose candidates without relaxation feedback~\cite{xie2022cdvae,zeni2025mattergen}. Such pipelines cannot recover when their initial proposals miss stable configurations~\cite{abolhasani2023rise,yang2025stable}; the GNoME project~\cite{merchant2023scaling} typifies both the promise and the ceiling of this paradigm, discovering 2.2~million crystal structures through GNN-guided screening yet lacking a per-configuration physical-feedback loop to diagnose and revise individual proposals that fail. Prioritizing among combinatorially many candidates instead requires \textit{chemical reasoning}---judging which binding motifs are plausible and when a predicted configuration is geometrically or chemically inconsistent---which fixed open-loop pipelines are structurally incapable of providing.

Large language models (LLMs) are natural candidates for this reasoning role, encoding broad chemical knowledge and performing complex compositional reasoning~\cite{mirza2024chembench,bran2024chemcrow,jablonka2023fourteen,ramos2025review}, as surveyed across materials discovery, property prediction, and autonomous experimentation~\cite{jiang2025applications,miret2025enabling}. Closed-loop platforms such as A-Lab~\cite{szymanski2023alab}, ChemCrow~\cite{bran2024chemcrow}, and Coscientist~\cite{boiko2023autonomous} have shown that feeding experimental or computational results back into the planning loop lets LLM agents coordinate multi-step chemical workflows and navigate search spaces that previously required expert chemists~\cite{zimmermann2025llm,yang2026quasar,stewart2026graphagents,chandrasekhar2026catalyst,weietal2025agentic}. Extending this success to adsorption-configuration search, however, remains underexplored. Adsorb-Agent~\cite{ock2026adsorbagent} took a pioneering step by coupling an LLM planner with an EquiformerV2~\cite{liao2024equiformerv2} MLFF to generate and enumerate adsorption hypotheses, but as an open-loop method it receives no feedback from physical relaxations and is never informed whether its proposals reached the intended binding motif or failed through structural instability. This single-pass regime produces failures that grow with surface complexity---unreliable initial site predictions, no interpretability of the underlying simulations, and systematic site biases that propagate unchecked and silently exclude stable configurations. These failure modes are intrinsic to open-loop search and motivate the remedy long validated by active learning~\cite{settles2012active} and Bayesian optimization~\cite{snoek2012practical}: iteratively feeding evaluation results back into the decision policy, which recent autonomous-discovery work likewise finds essential for reliable performance in complex search spaces~\cite{abolhasani2023rise,miret2025enabling}.

\begin{figure}[t]
    \centering
    \includegraphics[width=0.5\linewidth]{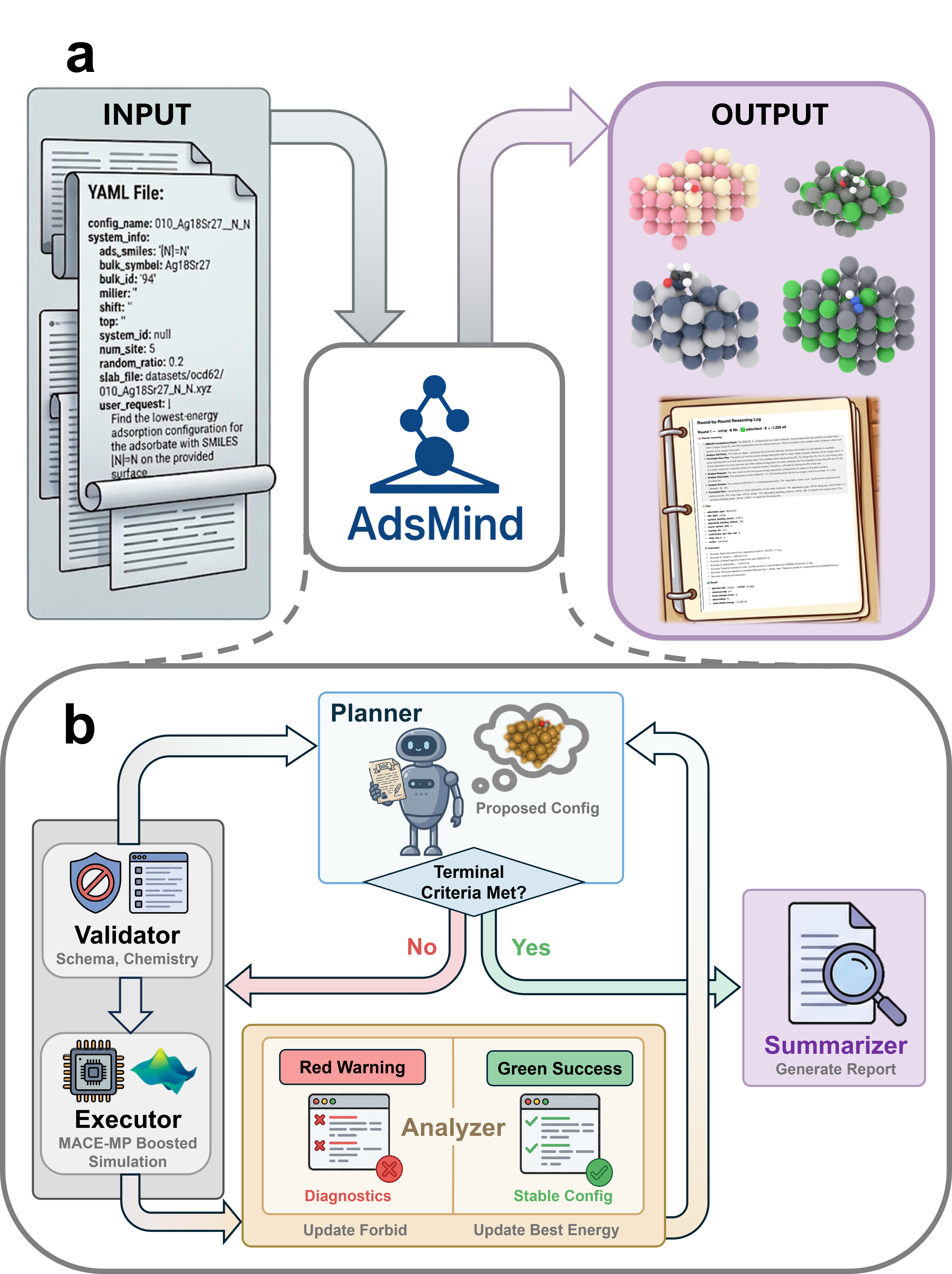}
    \caption{\textbf{Overview of AdsMind.} (a) Schematic diagram of the AdsMind, showing input format and test benchmark datasets: AA20 (20 surface-adsorbate cases) and OCD-GMAE62 (62 oxide, intermetallic, monometallic, and compound surfaces across two testing tiers).
    (b) Schematic diagram of the AdsMind closed-loop multi-agent framework, showing the Planner, Validator, Executor, Analyzer, and Summarizer agents.}
    \label{fig:pipeline}
\end{figure}

To overcome these limitations, we introduce \textbf{AdsMind} (\textbf{Ads}orption configuration discovery with \textbf{M}achine \textbf{in}telligence and relaxation fee\textbf{d}back), a closed-loop multi-agent framework in which each MLFF relaxation result feeds back into the LLM planner, letting the agent detect, diagnose, and recover from its own reasoning errors (Figure~\ref{fig:pipeline}). This design realizes three properties that open-loop agents cannot achieve simultaneously: reliable convergence through error recovery (enabled by FORBID constraints), self-reflection through iterative use of relaxation feedback, and interpretability through physically grounded feedback artifacts (enabled by Chemical Slip detection). AdsMind builds on our earlier prototype, AdsKRK, developed during the 2025 Large Language Model Hackathon for Applications in Materials Science and Chemistry~\cite{roy2026knowledgeactionoutcomes2025}.

We evaluate AdsMind across 82 surfaces (20 in AA20 and 62 in OCD-GMAE62~\cite{adsmt}) with four LLM backends and the MACE-MP-0 force field, benchmarking against conventional baselines and the open-loop agent Adsorb-Agent~\cite{ock2026adsorbagent}. DFT validation with VASP~\cite{kresse1996efficient} using the Perdew--Burke--Ernzerhof (PBE) exchange--correlation functional~\cite{perdew1996generalized} and projector augmented-wave (PAW) pseudopotentials on six representative AA20 systems reveals that the open-loop baseline produces qualitative sign errors on molecular adsorbates, whereas AdsMind preserves 100\% sign-correctness with competitive quantitative agreement. Systematic comparison with broader sampling strategies further delineates the reliability--depth--relaxation-budget trade-off, defining the operating envelope in which closed-loop refinement adds value. By delivering reliability, self-reflection, and interpretability simultaneously, AdsMind establishes a foundation for autonomous adsorption-configuration search.

\section{Methods}
\label{sec:methods}
We detail the five specialized modules of AdsMind---Planner, Validator, Executor, Analyzer, and Summarizer---and the implementation of its three closed-loop mechanisms: \textit{Chemical Slip} detection, the \textit{FORBID} directive, and the \textit{TERMINATE} directive.

\subsection{Closed-Loop Agent Architecture}

\subsubsection{Planner and Validator: Adsorption Hypothesis Generation}

The \textbf{Planner} serves as the entry point of each search iteration. Given a user request consisting of an adsorbate SMILES string, a relative path to a surface slab structure file, available surface-site description, and the running history of earlier attempts in the same search (defined below), it proposes a structured adsorption hypothesis in JSON format. Each hypothesis specifies the canonical site type (\textit{ontop}, \textit{bridge}, or \textit{hollow}), the intended surface binding atomic symbols, and the adsorbate binding indices~\cite{norskov2011density}. This fixed vocabulary makes every hypothesis checkable before simulation but cannot express every adsorption motif, as discussed under \textit{Limitations} in the Discussion.

The running history is the record of the preceding attempts in the same search, one entry per attempt. Each entry restates the planned hypothesis and reports the outcome of its relaxation: whether the adsorbate stayed intact, rearranged, or dissociated, or whether the calculation failed; the relaxed site type and the surface atoms bonded to the adsorbate; any \textit{Chemical Slip} warning or convergence tag; and the adsorption energy. Because each search is limited to five attempts, the history holds at most five entries, and in all benchmark runs it remained below 1800 characters. Hypotheses rejected by the \textbf{Validator} are returned to the \textbf{Planner} as a separate error message and count toward the same five-attempt budget. A complete example is reproduced in the Supporting Information.

However, LLMs are imperfect single-pass predictors of binding modes and site types. Under the \textit{1-Shot} protocol, 56--72\% of valid non-dissociated placements on AA20 and 57--72\% on OCD-GMAE62, depending on the backend, relax away from the planned surface-binding motif (Figure~\ref{fig:AA20_convergence}b and Table~S14). Pooled over the four backends, the fractions are 65.8\% on AA20 and 66.2\% on OCD-GMAE62. Many of these slips keep the planned surface elements and change only the number of bonded surface atoms (33 of the 48 slips on AA20). To let the \textbf{Planner} revise its hypotheses in response to physical evidence, namely the relaxed structures and the diagnostics that the \textbf{Analyzer} derives from them, its prompt incorporates the running history together with feedback signals derived from three closed-loop mechanisms.

First, \textit{Chemical Slip} detection compares the sorted element symbols of the surface atoms that the hypothesis intended to bind with those of the surface atoms bonded to the adsorbate anchor after relaxation, pooling both anchors for a two-point hypothesis, and flags a slip whenever the two lists differ in element or number. No slip is flagged when the adsorbate is left with no bonded surface atom. A change of site type, such as \textit{ontop} to \textit{hollow}, therefore counts as a slip even on a single-element surface. A slip does not reject the relaxed configuration, which keeps its energy, stays in the history, and remains eligible as the reported best result. What changes is the next prompt, which tells the \textbf{Planner} that the planned site proved unstable under the MLFF and should be treated as invalid. For example, an adsorbate planned on a bridge site that relaxes into a neighboring hollow site is flagged, even though the hollow site may be the physically preferred one. Such outcomes are frequent. The best configuration reported by the \textit{Full} workflow comes from a slipped attempt in 61.2\% of runs on AA20 and 69.4\% on OCD-GMAE62 (Table~S14). Second, after a slip, the \textit{FORBID} directive instructs the \textbf{Planner} not to plan the same surface-element combination again unless it has a strong geometric reason to expect a different environment to stabilize it. The directive acts through the prompt, whereas the \textbf{Validator} rejects only exact repeats of earlier hypotheses. Because a single slip can discourage a surface-element combination that would be favorable in another environment, \textit{FORBID} may narrow the search. The \textit{w/o Forbid} ablation measures this effect (Tables~S9 and~S11), and the resulting trade-off is discussed under \textit{Limitations}. Third, the \textit{TERMINATE} directive uses convergence signals from the search history to enable early stopping. Disabling all three mechanisms and restricting the search to a single attempt yields the 
\textit{1-Shot} baseline, corresponding to an open-loop search protocol with respect to post-relaxation feedback.

Even with iterative feedback, the \textbf{Planner} may still generate hypotheses that are geometrically or chemically malformed at the schema level. Such broken hypotheses can waste computationally expensive structural relaxation budget and introduce non-physical entries into the running memory. To prevent this, the \textbf{Validator} enforces three constraints before any physical simulation is launched: (1) the requested binding indices must be valid and correspond to heavy atoms, unless the adsorbate contains no heavy atoms, such as atomic H; (2) the adsorbate binding indices and surface binding atom list must be compatible with the planned site type: \textit{ontop} uses one adsorbate binding index and one surface atom, \textit{bridge} permits one or two adsorbate binding indices and one or two surface atoms, and \textit{hollow} permits one or two adsorbate binding indices with at least three surface atoms; and (3) the same hypothesis must not have been evaluated previously, thereby preventing redundant calculations. Hypotheses that fail validation receive immediate validation-failure feedback, prompting the \textbf{Planner} to revise its proposal without advancing to physical simulation. Table~S1 lists the complete set of checks.

\subsubsection{Executor and Analyzer: Physical Simulation and Scientific Analysis}

The validated hypothesis produced by the \textbf{Planner} is symbolic: it specifies chemical symbols, adsorption site types, and binding indices, but not explicit atomic coordinates. The \textbf{Executor} converts this symbolic hypothesis to a concrete adsorbate--slab structure and performs structural relaxation used to assess whether the proposed adsorption mode is physically plausible.

Before adsorbate placement, the input slab is standardized by rebuilding a clean ASE slab object from the input symbols, positions, cell, and periodic boundary conditions. If the in-plane cell vectors are shorter than 6~\AA, the slab is expanded along the corresponding in-plane direction by a factor of two, and the vacuum thickness is checked against a 15~\AA\ threshold. A surrogate-SMILES geometry layer first maps the specified adsorbate binding atoms to proxy markers and constructs initial placements for \textit{ontop}, \textit{bridge}, and \textit{hollow} adsorption modes. For each hypothesis, the \textit{Executor} samples a capped set of candidate surface sites and adsorbate conformers, using up to eight filtered surface sites and up to four conformers per site by default. The top-ranked candidate identified by a single-point energy screening is then selected for full structural relaxation.

A two-step evaluation procedure is used. Candidate structures are first evaluated using rapid MACE-MP-0 single-point calculations to discard geometrically pathological structures before consuming the full relaxation budget. In the primary benchmark, Langevin warmup is disabled (\texttt{md\_steps}=0) under the frozen MACE-MP-0 (small) protocol. The best surviving candidate is then relaxed using the BFGS algorithm~\cite{nocedal2006numerical} with a force convergence threshold of $f_{\max}=0.10$~eV/\AA\ and a maximum of 200 optimization steps. Atoms in the bottom one-third of the slab, identified by their $z$ coordinates, are fixed during structural relaxation, whereas the upper slab layers and adsorbate atoms are fully relaxed.

The adsorption energy is defined as 
\begin{equation}
    E_{\text{ads}} = E_{\text{total}} - E_{\text{surface}} - E_{\text{adsorbate}},
\end{equation}
where $E_{\text{total}}$ is the relaxed adsorbate--slab energy, $E_{\text{surface}}$ is the single-point energy of the corresponding bare surface, and $E_{\text{adsorbate}}$ is the energy of the relaxed isolated adsorbate. Complete default parameter settings are provided in the Supporting Information (Table~S3).

The \textbf{Executor} generates relaxed trajectories, final atomic structures, and energy records, which are not directly interpretable by the LLM-based \textbf{Planner}. The \textbf{Analyzer} bridges this semantic gap by parsing the relaxed trajectory into chemically meaningful diagnostics and writing them into the running memory for the subsequent iterations.

Connectivity is determined from an adjacency matrix $A$, where atoms $i$ and $j$ are considered bonded if 
\begin{equation}
    d_{ij} \le \gamma (R_i + R_j),
\end{equation}
where $d_{ij}$ is the interatomic distance and $R_i$ and $R_j$ are covalent radii. We use $\gamma = 1.35$ for intramolecular dissociation checks and $\gamma = 1.30$ for surface--adsorbate coordination analysis, increased to $\gamma = 1.45$ when $E_{\text{ads}} < -0.5$~eV. Dissociation is flagged when the adsorbate subgraph contains more than one connected component. The coordination environment of the anchor atom or atoms is then classified as \textit{ontop}, \textit{bridge}, or \textit{hollow} based on their bonding pattern within the first coordination shell.

The \textbf{Analyzer} records slip events, adsorption energies, and bond-integrity diagnostics in the running memory. These diagnostics inform subsequent \textbf{Planner} proposals through the \textit{Chemical Slip} feedback and the \textit{FORBID} directive, thereby converting relaxation outcomes into chemically grounded search constraints.

\subsubsection{Summarizer: Terminal Reporting and Early-Stopping Handling}

The \textbf{Summarizer} handles terminal reporting, while convergence information is accumulated in the running history during the Executor--Analyzer loop. After each physical attempt, the Analyzer returns diagnostics that are converted into a structured history entry containing the adsorption energy, planned and relaxed site types, chemical-slip status, bond-integrity diagnostics, and dissociation flag. When a relaxed configuration has the same site fingerprint as the current best result and differs by less than 0.05~eV, the history is tagged as converged to the known best configuration; energy-degenerate but geometrically distinct configurations are explicitly retained as separate discoveries. The \textbf{Planner} reads these tags in the next iteration. If the \textit{TERMINATE} mechanism is enabled, the Planner may emit \texttt{"action": "terminate"}, after which the Validator routes the workflow directly to the Summarizer. The workflow also terminates when the maximum attempt budget or validation-retry budget is exhausted.

Upon terminal routing, the \textbf{Summarizer} produces a structured Markdown report from the recorded run state. The report documents the run context, including the surface file, adsorbate, SMILES string, LLM backend, calculator backend, random seed, relaxation mode, and attempt budget; the best molecular adsorption energy and corresponding configuration; validation failures; a per-iteration table of planned versus relaxed binding sites, slip events, and dissociation flags; per-round Planner reasoning and plan JSON; and a final narrative analysis grounded in the recorded diagnostics. If a lower-energy dissociated state was observed, it is retained as a thermodynamic warning rather than replacing the reported molecular adsorption configuration. The report includes two deterministic visualizations generated outside the LLM reasoning loop: a rendering of the best configuration and an iteration-by-iteration energy convergence curve. Visualization failures are logged in the report and do not affect the selected configuration, energies, or search decisions.

\subsection{Experimental Setup and Ablation Variants}

We evaluate AdsMind under a frozen experimental configuration using the MACE-MP-0 (small) foundation model~\cite{batatia2022mace,batatia2023foundation} as the primary MLFF calculator, configured to run on CPU with float32 precision and dispersion disabled. This configuration is held fixed across all primary AdsMind ablations and MACE-MP-0-based baseline comparisons, thereby minimizing confounding from the choice of potential-energy model and isolating the effect of the search strategy. Separate sensitivity checks probing MLFF calculator settings use the MACE-MP-0 (large) variant, configured to run on CUDA with float64 precision and dispersion enabled. Calculator configurations and default execution parameters are provided in the Supporting Information (Tables~S2 and~S3).

The three closed-loop mechanisms described above define the primary ablation variants. The \textit{Full} configuration enables all three mechanisms. The \textit{1-Shot} baseline disables all three mechanisms and restricts the \textbf{Planner} to a single attempt, yielding the open-loop baseline against which iterative correction is measured. The \textit{w/o Term} variant disables only the \textit{TERMINATE} directive, retaining \textit{Chemical Slip} feedback and the \textit{FORBID} directive while removing convergence-based early stopping. In this setting, the run proceeds until the fixed attempt budget is exhausted. The \textit{w/o Slip} variant disables \textit{Chemical Slip} feedback and its associated \textit{FORBID} updates while retaining the termination mechanism. The \textit{w/o Forbid} variant retains \textit{Chemical Slip} feedback but disables the explicit \textit{FORBID} directive.

All benchmark runs use four reasoning backends with sampling temperature set to 0 where supported: Gemini-2.5-Pro~\cite{gemini2026modelcard}, GPT-5.4~\cite{openai2026gpt5}, Claude-Sonnet-4.6~\cite{anthropic2026claude}, and Grok-4~\cite{grok2025blog}. Natural chemical failures, including dissociation or rearrangement during structural relaxation, are retained in the success-rate denominator. Energy-based comparisons are computed over runs with valid molecular adsorption energies. Runs without a valid molecular product are not assigned artificial energies. Natural dissociation or rearrangement events are counted as unsuccessful attempts in success-rate calculations, whereas external service errors and calculator-level failures are reported separately when present.

\subsection{Non-LLM Baseline}

To isolate the contribution of the LLM-driven search strategy from that of the underlying structural relaxation model, we implement a non-LLM heuristic control using the same MACE-MP-0 (small) MLFF calculator. This baseline identifies high-symmetry surface sites using the AutoAdsorbate enumeration method~\cite{fako2025simple}, yielding a median of 56 sites per case on AA20 (range 25--98) and 61.5 sites per case on OCD-GMAE62 (range 19--126). Each generated candidate is relaxed using the same BFGS relaxation settings as the AdsMind \textbf{Executor} ($f_{\max}=0.10$~eV/\AA, 200 steps, bottom one-third of the surface slab fixed).

The heuristic baseline operates in an open-loop enumeration regime and does not use the post-relaxation chemical diagnostics, running memory, or corrective feedback mechanisms applied by AdsMind. This design makes it a representative conventional enumeration baseline. Raw relaxed structures and energies are retained for downstream chemical-validity audit, allowing chemically unstable outcomes such as dissociation or rearrangement during structural relaxation to be identified rather than corrected during the search.

\subsection{DFT Reference Calculations}\label{sec:dft_reference}

To validate the physical fidelity of MLFF-relaxed configurations against a first-principles reference, we performed density functional theory (DFT) calculations on six representative AA20 systems. These systems were selected to cover three representative surface models---including Mo$_3$Pd(111), CuPd$_3$(111), and Pt(111)/(100)---and three adsorbates, H, NNH, and OH. The specific surface--adsorbate pairings are listed in the Supporting Information (Table~S5). The selected cases include both simple and structurally complex adsorbates, where MLFF-based search strategies can exhibit the largest quantitative deviations from DFT.

The DFT adsorption energy was evaluated using the same definition as the MLFF workflow,
\begin{equation}
    E_{\mathrm{ads}} = E_{\mathrm{slab+adsorbate}} - E_{\mathrm{slab}} - E_{\mathrm{adsorbate}},
\end{equation}
where all three terms are final self-consistent field (SCF) total energies. Isolated adsorbates were separately optimized and then evaluated by final SCF calculations using the optimized gas-phase structures, with cell and $k$-point settings generated consistently using the same VASPKIT workflow.

All DFT calculations were performed using the Vienna Ab initio Simulation Package (VASP) 6.4.2~\cite{kresse1996efficient,kresse1999ultrasoft} with the PBE exchange--correlation functional~\cite{perdew1996generalized} and projector-augmented-wave (PAW) pseudopotentials. Input files were generated using VASPKIT~1.5.1~\cite{wang2021vaspkit}.

Atomic models were constructed using Materials Studio 2023. Slab models consisted of four atomic layers expanded to $2\times2$ supercells with 15~\AA\ of vacuum perpendicular to the surface. The bottom two layers were fixed to mimic the underlying bulk, whereas the upper layers and adsorbate atoms were allowed to relax. A plane-wave cutoff of 400~eV was used for pristine slabs, whereas 520~eV was used for adsorbate-containing systems and isolated adsorbates, corresponding to approximately 1.3 times the largest ENMAX value among the elements present.

Pristine slabs were treated as non-spin-polarized systems (ISPIN~=~1). Spin polarization was enabled for all adsorbate-containing calculations (ISPIN~=~2) with initial magnetic moments assigned as follows: H adsorption, 1~$\mu_\mathrm{B}$ on the H atom and 0 on all substrate atoms; OH adsorption, 1~$\mu_\mathrm{B}$ on the H atom and 0 on all other atoms; and NNH adsorption, 1~$\mu_\mathrm{B}$ on the first N atom in the NNH model and 0 on all other atoms. These spin settings were used for both structural relaxations and subsequent SCF calculations.

Structural relaxations were performed for both pristine slabs and adsorbate-covered slabs, with the latter constructed from fully relaxed pristine slabs. The Methfessel--Paxton smearing scheme~\cite{methfessel1989high} was used with ISMEAR~=~1 and $\sigma=0.2$~eV. Monkhorst--Pack $k$-point sampling~\cite{monkhorst1976special} with a reciprocal-space density of 0.03~\AA$^{-1}$ was used for structural relaxation, and $\Gamma$-centered meshes with a spacing of 0.02~\AA$^{-1}$ were used for final SCF calculations. All $k$-point meshes were generated using VASPKIT. The maximum numbers of ionic and electronic steps were both set to 200, and ALGO~=~Normal was used. Ionic relaxations were converged to a force threshold of $10^{-2}$~eV/\AA, and the electronic convergence threshold during relaxation was $10^{-5}$~eV.

After relaxation, the WAVECAR and CHGCAR files were used to initialize final SCF calculations, for which the electronic convergence criterion was tightened to $10^{-7}$~eV while all other parameters, including cutoff energy, spin polarization, ISMEAR, and ALGO, were kept identical to those used during relaxation.

No dipole correction or empirical dispersion correction was applied in the production DFT calculations. Test calculations with dipole correction and Grimme dispersion corrections (IVDW~=~11/12) on representative systems changed adsorption energies by no more than $\sim10^{-2}$~eV while substantially increasing computational cost. These corrections were therefore omitted from the production calculations.

\subsection{Evaluation Datasets}

The primary benchmark, \textbf{Adsorb-Agent20} (AA20), is based on the 20-case manifest introduced in Adsorb-Agent~\cite{ock2026adsorbagent} and its open-source implementation~\cite{adsorbagent}, spanning the adsorbates H, OH, NNH, CH$_2$CH$_2$OH, OCHCH$_3$, and ONN(CH$_3$)$_2$ on intermetallic and monometallic surfaces. The complete five-configuration $\times$ four-backend matrix was executed on all 20 cases, yielding 400 benchmark runs. The three main configurations, \textit{1-Shot}, \textit{w/o Term}, and \textit{Full}, are reported in the main text, whereas the remaining two ablations, \textit{w/o Slip} and \textit{w/o Forbid}, are provided in the Supporting Information. Natural chemical failures are counted as unsuccessful attempts in success-rate calculations. Energy-based metrics are computed over runs with valid molecular adsorption energies. Runs without a valid molecular product are reported as failures or missing entries rather than assigned artificial energies. To enable multi-objective comparison across backends, Energy Accuracy, Backend Robustness, and Cross-LLM Agreement are normalized such that higher values indicate better performance. Their definitions are provided in the Supporting Information.

For validation beyond the monometallic and intermetallic scope of AA20, we assembled \textbf{OCD-GMAE62} from the OCD-GMAE benchmark introduced by AdsMT~\cite{adsmt}, a 62-case dataset spanning intermetallic, monometallic, and chemically diverse compound surfaces. This dataset evaluates whether closed-loop reasoning remains effective across more diverse surface chemistries. We employ a two-tier evaluation protocol. Tier~1 runs the complete five-configuration $\times$ four-backend matrix across all 62 cases, yielding 1240 benchmark runs. The three main configurations are reported in the main text, and the additional two ablations are provided in the Supporting Information. This tier quantifies the performance penalty of the \textit{1-Shot} baseline, the effects of individual ablations, and the degree of cross-backend agreement.

Tier~2 is a stability audit that repeats 12~subsampled cases with $N=3$ repeated runs for each case, variant, and backend under the same four-backend protocol. The 12-case subset and run-selection protocol are provided in the Supporting Information. For each selected case, variant, and backend, the reported $N=3$ repeats retain run~1, corresponding to the default random seed of 42, and add the two runs from an initial $N=5$ repeat set that minimize the energy range. Tier~2 quantifies run-to-run stochasticity under fixed runtime settings and complements the Tier~1 cross-backend analysis by contextualizing whether observed variance arises primarily from repeated sampling or from backend-level differences. All experiments use the fixed MACE-MP-0 (small) CPU protocol.

\section{Results}

\subsection{DFT Verification and Comparison}

To assess the physical fidelity of adsorption energies obtained from LLM-driven adsorption-search workflows, we compared AdsMind with the previously reported EquiformerV2-based Adsorb-Agent results from Ref.~\citenum{ock2026adsorbagent} against DFT/PBE reference calculations performed with VASP (Methods section). The comparison was conducted on six representative AA20 systems (cases~1, 2, 3, 4, 9, and~10), covering three surface models---Mo$_3$Pd(111), CuPd$_3$(111), and Pt(111)/(100)---and three adsorbates, H, NNH, and OH. These systems include both simple atomic adsorption and molecular reaction intermediates on catalytically relevant metal surfaces. Ordered bimetallic surfaces such as Mo$_3$Pd and CuPd$_3$ are relevant to nitrogen-reduction catalysis~\cite{zhou2023enhanced}, whereas Pt serves as a standard noble-metal reference for hydrogen electrocatalysis~\cite{cheng2016platinum}. Because the reported Adsorb-Agent values are based on EquiformerV2 whereas AdsMind uses MACE-MP-0 (small), this comparison should be interpreted as a DFT reference check of the reported workflow outputs rather than as a force-field-controlled architecture-only ablation.

\begin{figure}[t]
\centering
\includegraphics[width=0.95\linewidth]{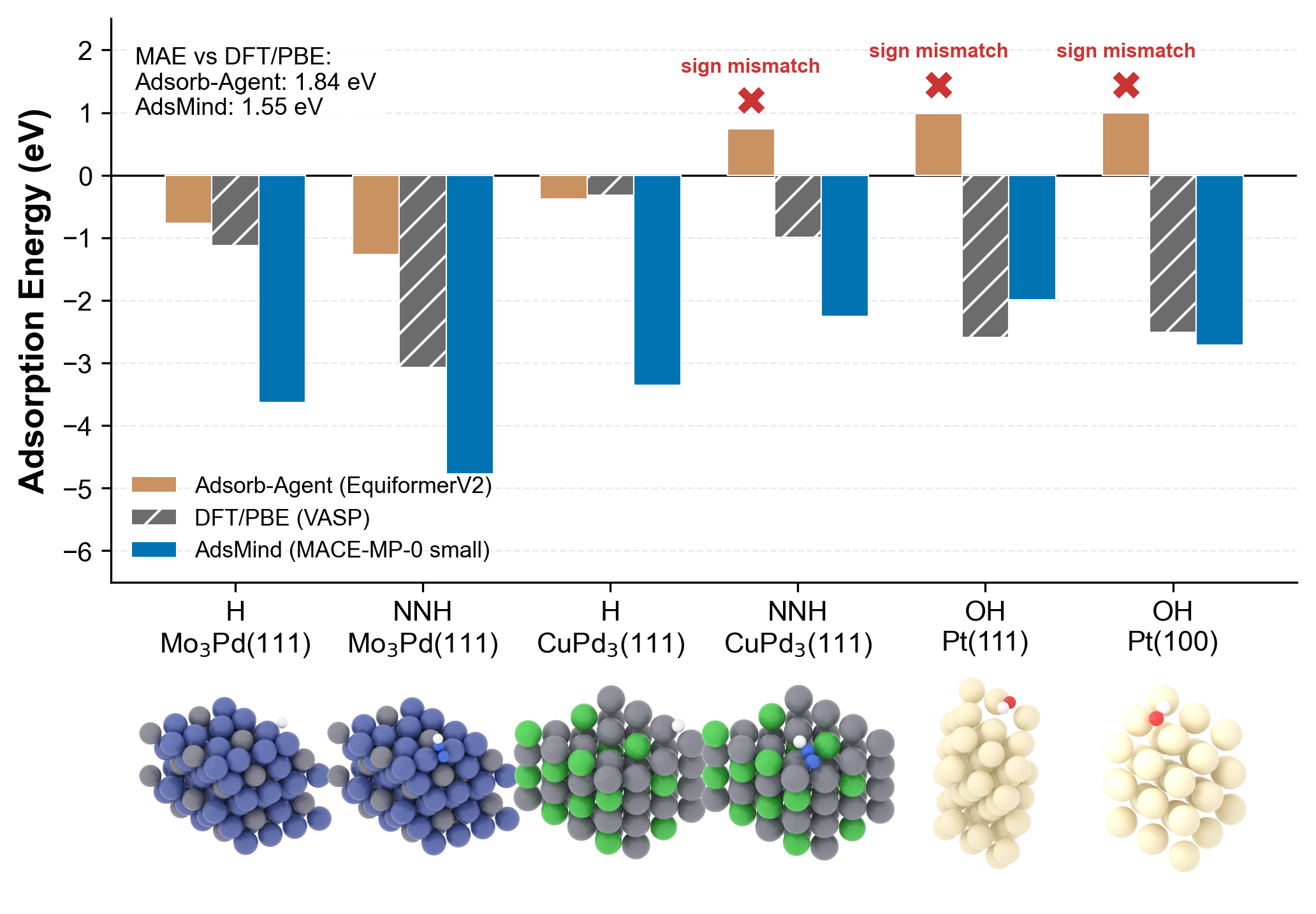}
\caption{\textbf{DFT reference comparison on six representative AA20 systems.} Grouped bar chart comparing adsorption energies (eV) from the Adsorb-Agent EquiformerV2 results reported in Ref.~\cite{ock2026adsorbagent}, DFT/PBE reference calculations with VASP, and AdsMind [MACE-MP-0 (small), GPT-5.4 \textit{Full}]. AdsMind preserves the adsorption-energy sign relative to PBE across all six tested systems (MAE 1.55~eV), whereas the reported Adsorb-Agent adsorption-energy values exhibit sign mismatches for CuPd$_3$--NNH and OH on Pt(111)/Pt(100) (MAE 1.84~eV).}
\label{fig:vasp_validation}
\end{figure}

Figure~\ref{fig:vasp_validation} compares the reported Adsorb-Agent values and AdsMind outputs with the DFT/PBE reference calculations. The Adsorb-Agent adsorption-energy values show substantial deviations from PBE, with a mean absolute error (MAE) of 1.84~eV and a maximum deviation exceeding 3.5~eV for the Pt(111)--OH pair. The most consequential discrepancies are qualitative sign mismatches for molecular adsorbates. Adsorb-Agent predicts endothermic binding for OH on both Pt(111) and Pt(100) surfaces ($+0.99$~eV for both), whereas PBE yields strongly exothermic adsorption energies ($-2.59$~eV and $-2.51$~eV, respectively). Similarly, NNH on the CuPd$_3$(111) surface is predicted as endothermic by Adsorb-Agent ($+0.745$~eV), whereas the PBE reference is exothermic ($-0.996$~eV). If adsorption-energy sign were used as a screening criterion, such sign mismatches could lead to false-negative decisions for otherwise viable surface--adsorbate pairs, such as Pt--OH and CuPd$_3$--NNH. In contrast, for the simple atomic hydrogen adsorption cases, Adsorb-Agent remains relatively close to PBE, with a maximum error $\le 0.36$~eV. This pattern suggests that the largest qualitative failures in this validation set are associated with molecular adsorbates and more complex binding geometries rather than simple atomic adsorption.

AdsMind yields a lower MAE against the PBE reference in this AA20 subset (1.55~eV) and, more importantly, shows no adsorption-energy sign mismatch across the six tested systems. Although AdsMind quantitatively overbinds some adsorption configurations on Mo$_3$Pd and CuPd$_3$ surfaces, consistent with broader evidence that universal pretrained ML potentials can show alloy-dependent errors~\cite{casillas2024evaluating}, it preserves the exothermic adsorption-energy sign for the Pt--OH and CuPd$_3$--NNH systems where the reported Adsorb-Agent adsorption-energy values show sign mismatches. The residual quantitative error is partly attributable to the MACE-MP-0 (small) MLFF backend. A sensitivity analysis comparing MACE-MP-0 (small) and MACE-MP-0 (large) shows that mean absolute energy shifts exceed 1~eV across the AA20 benchmark (Supporting Information).

No sign mismatches occur for the simple H-on-metal systems in either set of workflow outputs, whereas the observed sign mismatches are concentrated in non-H adsorbate cases involving OH and NNH. This pattern suggests that qualitative adsorption-energy failures in this validation subset are associated with adsorbate identity and binding complexity. Because the reported Adsorb-Agent adsorption-energy values are based on EquiformerV2 whereas the AdsMind values are based on MACE-MP-0 (small), this comparison should not be interpreted as an architecture-only attribution. Rather, the result supports the narrower conclusion that, within this six-system DFT subset, AdsMind avoids the molecular-adsorbate sign mismatches observed in the reported open-loop reference outputs.

Overall, AdsMind produces no adsorption-energy sign mismatches relative to PBE across the six tested systems, whereas the reported Adsorb-Agent values show sign mismatches in three of the four non-H cases. Because adsorption-energy sign can affect catalyst-screening decisions, these results highlight the practical importance of DFT reference checking and physically grounded relaxation feedback for molecular adsorbates. However, the limited DFT subset and the use of different MLFF backends preclude broader claims of DFT-level quantitative accuracy or architecture-only attribution across the full benchmark.

\subsection{Performance on the AA20 Benchmark}

\begin{figure}
\centering
\includegraphics[width=0.95\linewidth]{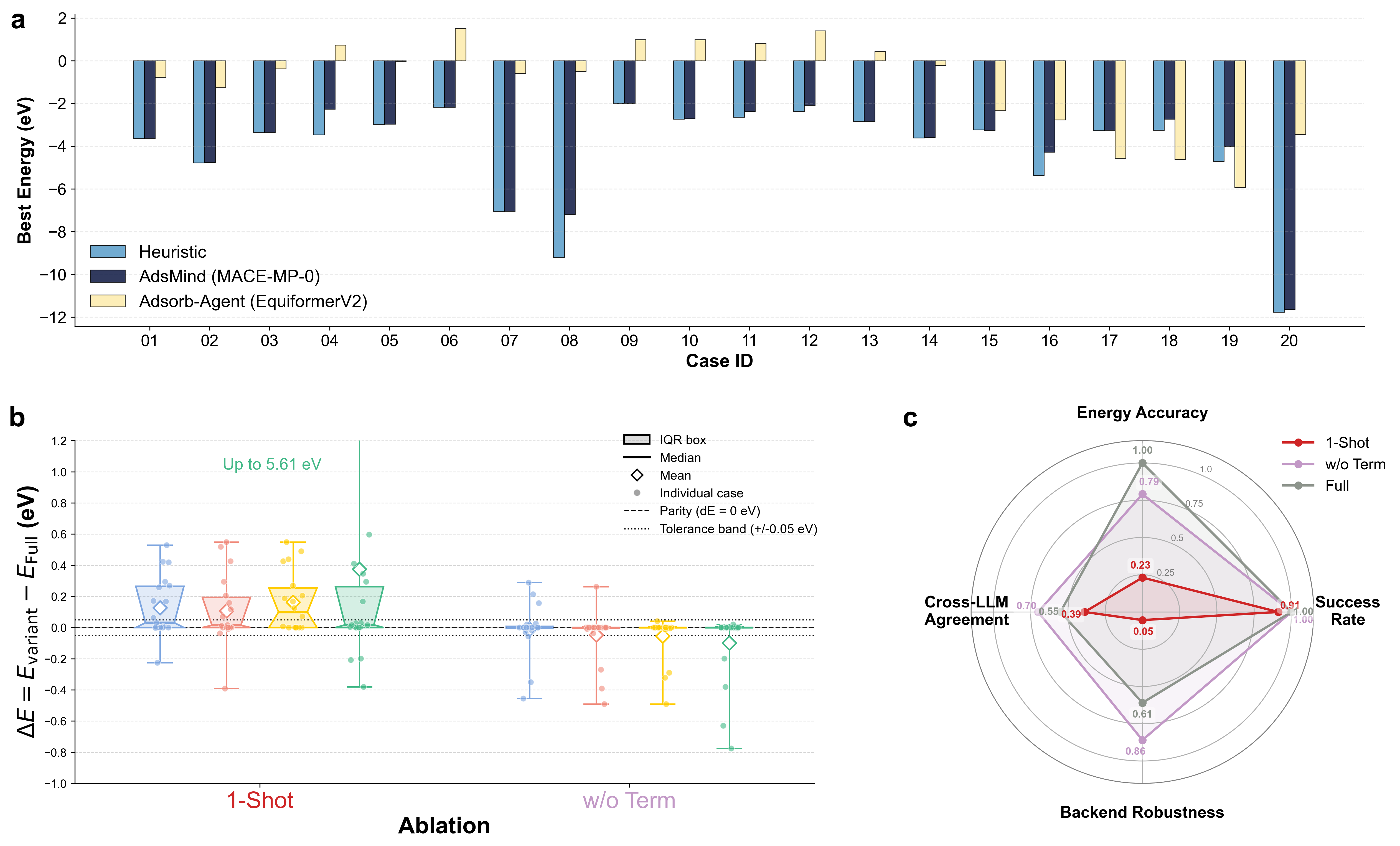}
\caption{\textbf{Performance and ablation analysis on the AA20 benchmark across four LLM backends.}
(a) Best-energy distributions comparing the reported Adsorb-Agent EquiformerV2 results, the heuristic enumeration baseline under the MACE-MP-0 (small) protocol, and \textit{AdsMind Full} with MACE-MP-0 (small) MLFF backend across the AA20 cases.
(b) Energy differences $\Delta E = E_{\mathrm{variant}} - E_{\mathrm{Full}}$ for the \textit{1-Shot} and \textit{w/o Term} variants, with dashed lines at $\Delta E = \pm0.05~\mathrm{eV}$ marking the tolerance band.
(c) Radar chart of \textit{Full}, \textit{1-Shot}, and \textit{w/o Term} across Success Rate, Energy Accuracy, Backend Robustness, and Cross-LLM Agreement. These normalized metrics are defined in the Supporting Information, with larger values indicating better performance.}
\label{fig:AA20_ablation}
\end{figure}

We first situate AdsMind within the landscape of existing adsorption-search strategies. On the 20-case AA20 benchmark, heuristic enumeration identifies more favorable raw MLFF adsorption energies than \textit{AdsMind Full} in 8 of 20 cases when using a 0.01~eV tolerance, reflecting its broader open-loop sampling of candidate adsorption sites. This broader search comes with a larger MLFF relaxation budget: the heuristic baseline enumerates a median of 56 candidate sites per case, with a range of 25--98. However, manual connectivity inspection of the saved top-three heuristic candidates indicates chemically unstable top-ranked outcomes in 25\% of cases (5/20), involving NNH dissociation or larger-adsorbate fragmentation in cases~04, 08, 16, 19, and~20. After filtering for chemical validity, the heuristic energy advantage remains in only 3 of 20 cases.

These results delineate a breadth--reliability trade-off in automated adsorption discovery. Broader open-loop sampling can identify more favorable raw MLFF adsorption energies, but it requires a substantially larger relaxation budget and can return chemically invalid low-energy structures. AdsMind occupies the reliability-first side of this trade-off, accepting some loss in raw energy depth in exchange for chemically audited convergence with a mean of 4.11 MLFF relaxations per case on AA20, more than an order of magnitude fewer than heuristic enumeration under the same MACE-MP-0 (small) structural relaxation protocol.

The ablation experiments quantify the contribution of the closed-loop feedback. Figure~\ref{fig:AA20_ablation}b shows the energy deviation of the \textit{1-Shot} and \textit{w/o Term} variants relative to \textit{Full} across the four LLM backends, with dashed $\Delta E=\pm0.05$~eV lines used as a practical comparison threshold. The \textit{1-Shot} configuration disables all three feedback mechanisms and restricts the Planner to a single attempt. It substantially underperforms \textit{Full}: only 47.9\% of valid \textit{1-Shot} runs fall within $|\Delta E| \le 0.05$~eV of the \textit{Full} result (35/73; 7 failed runs), compared with 78.7\% for \textit{w/o Term} (63/80). This tolerance-band comparison complements the mean $\Delta E$ analysis. Although extreme \textit{1-Shot} failures increase the mean penalty, the tolerance-band metric shows that \textit{1-Shot} often falls outside the \textit{Full} energy window across much of the distribution, whereas \textit{w/o Term} remains closer to \textit{Full} when \textit{Chemical Slip} feedback and \textit{FORBID} constraints are retained.

The mean per-case improvement of \textit{Full} over \textit{1-Shot} ranges from 0.108~eV for GPT-5.4 to 0.374~eV for Grok-4, with no backend showing a mean degradation. The magnitude of this gain varies across backends, consistent with differences in single-pass planning performance. GPT-5.4 shows the smallest mean improvement, suggesting less room for iterative recovery, but still gains 0.426~eV on case~02 and 0.549~eV on case~17. Grok-4 shows the largest mean gain, dominated by difficult adsorption cases such as case~20, where \textit{1-Shot} is 5.608~eV higher than \textit{Full}. Case~15 degrades under \textit{1-Shot} across all four LLM backends. This pattern is consistent with a feedback-based correction mechanism whose benefit depends on both backend planning behavior and adsorption-case difficulty.

The \textit{w/o Term} variant disables only convergence-based early stopping while retaining \textit{Chemical Slip} feedback and \textit{FORBID} constraints. It occasionally reaches marginally lower energies than \textit{Full}, with an overall mean deviation of $-0.052$~eV on AA20, reflecting the expected trade-off that a relaxed stopping criterion can permit deeper exploration while requiring additional MLFF structural relaxations.

Figure~\ref{fig:AA20_ablation}c summarizes the multi-objective comparison. \textit{Full} achieves a 100\% success rate and provides the most balanced operating point across reliability, energy accuracy, backend robustness, cross-LLM agreement, and relaxation-budget control. The \textit{w/o Term} variant can offer marginal improvements in raw energy depth and cross-backend agreement, but it does so by relaxing the early-stopping criterion and using additional MLFF relaxations. Overall, the radar chart indicates that \textit{Full} is not strictly optimal in every individual metric, but it delivers the most balanced reliability-first configuration among the main variants. Two additional variants that individually disable \textit{Chemical Slip} feedback or \textit{FORBID} constraints yield results close to \textit{Full} and are reported in the Supporting Information.

\begin{figure}
\centering
\includegraphics[width=0.95\linewidth]{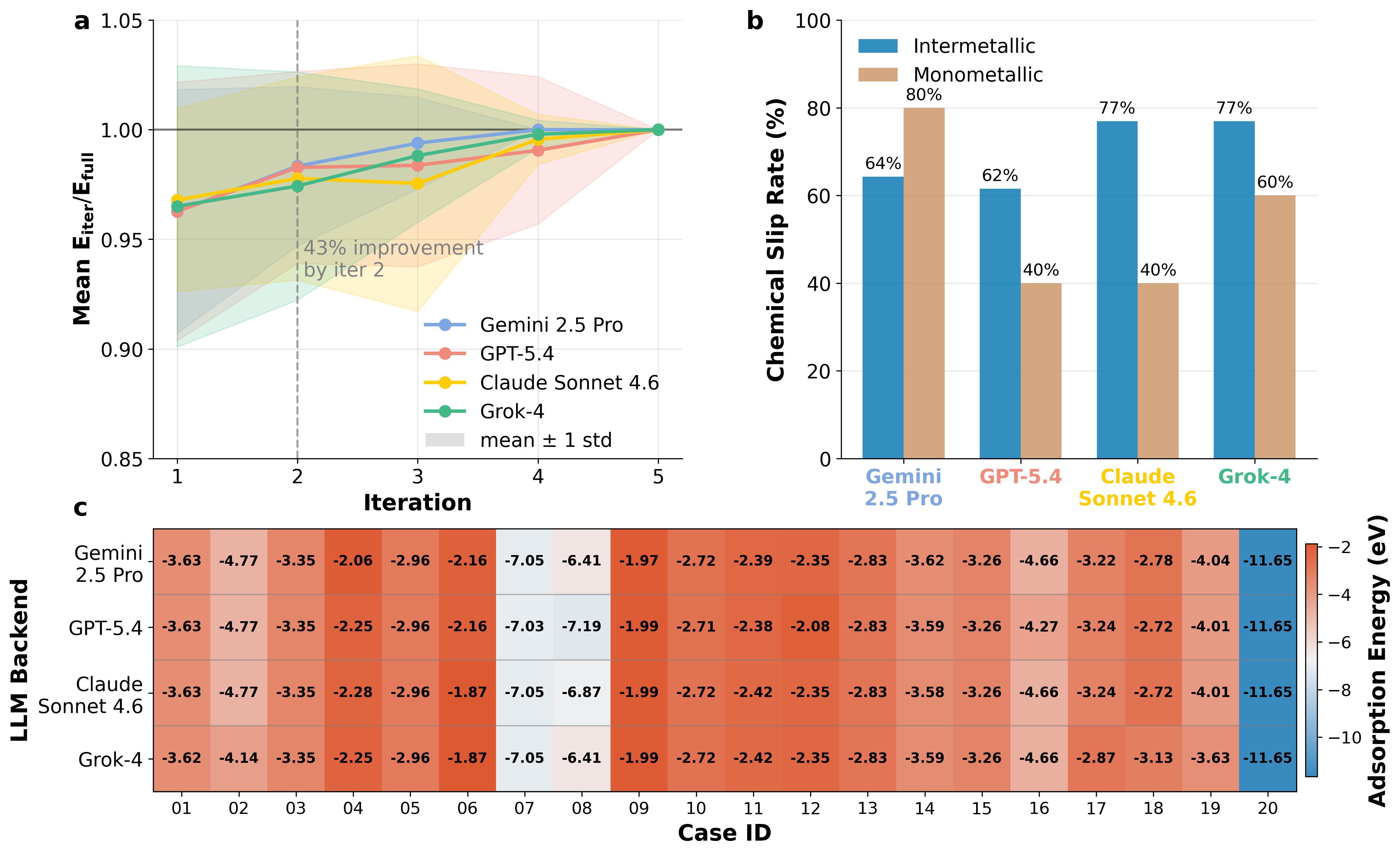}
\caption{\textbf{Advanced analysis on the AA20 benchmark.}
(a) Iteration-level convergence of the running-best adsorption energy relative to the final \textit{Full} result across 20 cases and four LLM backends, excluding dissociated attempts from running-best updates.
(b) \textit{Chemical Slip} analysis: \textit{1-Shot} slip rates among valid non-dissociated attempts, stratified by surface family (intermetallic vs.\ monometallic) across four LLM backends.
(c) Four-backend agreement heatmap showing the selected adsorption energy of the \textit{Full} variant across 20 cases and four LLM backends.}
\label{fig:AA20_convergence}
\end{figure}

The ablation results show that closed-loop feedback improves performance on the AA20 dataset. Figure~\ref{fig:AA20_convergence} further examines this improvement through iteration-level convergence, \textit{Chemical Slip} diagnostics, and cross-backend agreement.

First, the iteration-level convergence analysis (Figure~\ref{fig:AA20_convergence}a) shows that most of the closed-loop improvement is realized within the first few iterations. After excluding dissociated attempts from running-best updates, iteration~2 captures approximately 43\% of the initial gap to the final selected molecular adsorption energy, after which returns diminish across all four backends. This rapid convergence explains why the \textit{Full} configuration achieves its performance with a mean of only 4.11 MLFF relaxations per case on AA20 and supports the five-attempt budget as sufficient for stabilization in most AA20 cases.

Second, the prevalence of \textit{Chemical Slip} under \textit{1-Shot} conditions (Figure~\ref{fig:AA20_convergence}b) identifies a key failure mode of open-loop agents. Across four LLM backends, 61.5--76.9\% of valid non-dissociated \textit{1-Shot} attempts on intermetallic surfaces and 40.0--80.0\% on monometallic surfaces relax into a surface-binding motif that differs from the planned one, that is, undergo a \textit{Chemical Slip}. The elevated slip rates across all four LLM backends suggest that this behavior is not confined to a single LLM backend, but rather reflects the geometric sensitivity of adsorption-site placement. An open-loop agent, lacking post-relaxation feedback, may report the realized configuration without explicitly diagnosing whether it still matches the planned adsorption hypothesis, thereby silently conflating symbolic intent with physical outcome. The closed-loop architecture addresses this failure by feeding \textit{Chemical Slip} events back into the Planner context, allowing subsequent proposals to account for prior placement errors. The failure lies in leaving the slip unnoticed, not in the slip itself, which is often an ordinary relaxation outcome. On AA20, 69\% of the single-pass slips keep the planned surface elements and change only the number of surface atoms bonded to the anchor, most often from \textit{ontop} to \textit{hollow}, whereas on the compositionally more diverse OCD-GMAE62 surfaces 61\% change the bonded elements (Table~S14). The slipped configuration remains a candidate result: in the \textit{Full} runs, 61\% of the best configurations reported on AA20 and 69\% of those on OCD-GMAE62 come from attempts flagged as slips.

Third, the cross-backend heatmap (Figure~\ref{fig:AA20_convergence}c) shows that the \textit{Full} configuration reduces cross-backend dispersion in selected adsorption energies. Under \textit{Full}, the mean four-backend energy range across AA20 is 0.195~eV, with a median of 0.037~eV, and 7 of 20 cases agree within 0.01~eV. The largest disagreement occurs in case~08 (Ru$_3$Mo + NNH), where GPT-5.4 reaches $-7.19$~eV, Claude reaches $-6.87$~eV, and Gemini/Grok-4 cluster near $-6.41$~eV. Relative to the \textit{1-Shot} baseline, cross-backend agreement is substantially improved. On the 18 AA20 cases where \textit{1-Shot} succeeds on all four backends, the mean cross-backend range is 0.473~eV for \textit{1-Shot} and 0.158~eV for \textit{Full}. However, the agreement remains case-dependent rather than universal.

\subsection{Generalization to the OCD-GMAE62 Dataset}

To test whether the AA20 conclusions generalize beyond the predominantly monometallic and intermetallic scope of AA20, we evaluated AdsMind on OCD-GMAE62, a 62-case dataset spanning intermetallics (e.g., Hf$_{16}$Zn$_{48}$), chalcogenides (e.g., Mo$_{18}$S$_{72}$W$_{18}$), and other multi-component compound and alloy surfaces (e.g., Cd$_{24}$Pd$_{24}$, As$_{48}$Hf$_{48}$Ni$_{48}$), with adsorbates ranging from single atoms (H and N) to multi-atom species (e.g., C(C)O and C(=O)C). We employed a two-tier protocol: Tier~1 runs the complete variant~$\times$~backend matrix on all 62 cases, whereas Tier~2 repeats 12 subsampled cases with $N=3$ selected repeated runs for each variant and each of the four LLM backends to quantify run-to-run stochasticity under fixed runtime settings.

\begin{figure}[t]
\centering
\includegraphics[width=0.95\linewidth]{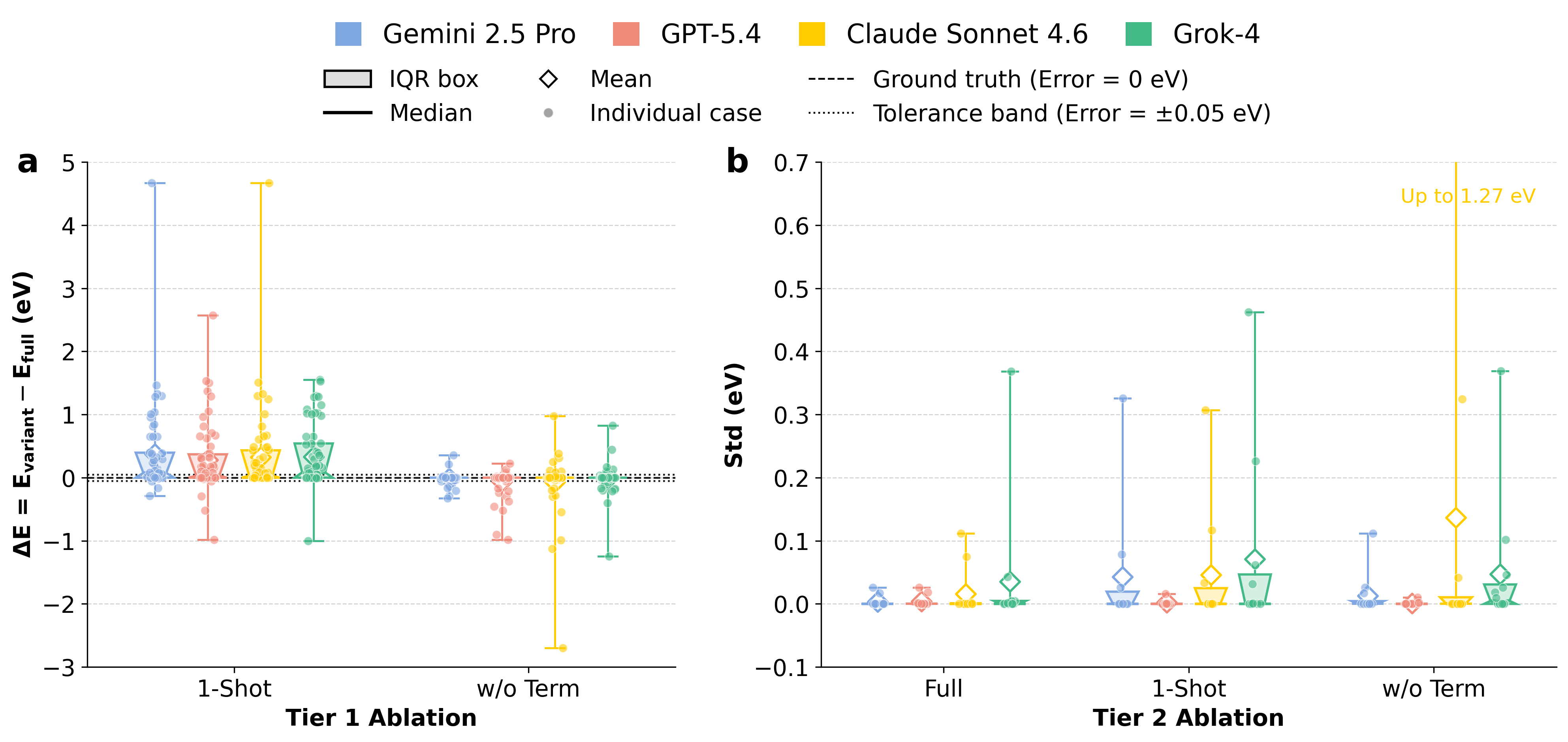}
\caption{\textbf{Two-tier evaluation overview on OCD-GMAE62.}
(a) Energy differences $\Delta E = E_{\mathrm{variant}} - E_{\mathrm{Full}}$ for the \textit{1-Shot} and \textit{w/o Term} variants, with dashed lines at $\Delta E = \pm0.05~\mathrm{eV}$ marking the tolerance band.
(b) Run-to-run standard deviation across the 12 Tier~2 cases, each evaluated with $N=3$ repeated runs for each setting.}
\label{fig:ocd62-two-tier}
\end{figure}

\subsubsection{Tier~1: Full-matrix evaluation}

\textit{Full} achieves a 98.8\% success rate on Tier~1 (245/248 successful runs), whereas \textit{1-Shot} drops to 89.5\% (222/248). The aggregate \textit{1-Shot} penalty is larger on OCD-GMAE62 than on AA20. The mean paired adsorption-energy difference, $\Delta E = E_{\mathrm{1-Shot}} - E_{\mathrm{Full}}$, is $+0.329$~eV (median 0.067~eV), compared with $+0.192$~eV on AA20. The $\pm0.05$~eV tolerance band (Figure~\ref{fig:ocd62-two-tier}a) provides a complementary metric. Only 44.6\% of successful same-backend paired \textit{1-Shot} comparisons (99/222) fall within the tolerance region, compared with 78.0\% of successful same-backend paired \textit{w/o Term} comparisons (191/245). This closely mirrors the pattern observed on AA20 (47.9\% vs.\ 78.7\%), suggesting that the performance penalty reflects a recurring effect of removing closed-loop feedback rather than an artifact of a single benchmark.

The three \textit{Full} failures are concentrated in a single case, case~053 (K$_{20}$ + C([CH$_2$])O), where three backends undergo natural adsorbate dissociation without retaining a valid molecular adsorption configuration, whereas the Claude backend later finds a non-dissociated O-bound configuration after switching the proposed binding atom. This pattern indicates that the observed \textit{Full} failures in Tier~1 arise from natural chemical events rather than framework-level execution errors. The aggregate mean $\Delta E$, however, conceals substantial case-level heterogeneity in the benefit of closed-loop feedback. In chemically complex cases involving multi-element surfaces or challenging adsorbates, iterative feedback provides large paired gains over \textit{1-Shot}: case~004 (Pt$_{16}$V$_{48}$ + [N]=O) improves by 0.95~eV on average across LLM backends, case~005 (As$_{48}$Hf$_{48}$Ni$_{48}$ + [NH$_2$]) by 0.88~eV, and case~021 (Al$_{20}$Au$_{20}$Y$_{20}$ + [NH$_2$]) by 0.79~eV.

In some cases where a single binding motif appears to dominate, iteration can add little beyond reliability guarantees: three cases yield identical mean energies for \textit{Full} and \textit{1-Shot} among successful same-backend paired comparisons. A backend-specific edge case occurs for the GPT-5.4 backend on case~008 (Ga$_{12}$Pd$_{36}$ + [NH]), where the Planner's first hypothesis reaches a low-energy basin ($-8.47$~eV), but subsequent feedback-constrained search in the \textit{Full} workflow converges to a shallower minimum ($-7.48$~eV, 0.99~eV worse than \textit{1-Shot}). This behavior is not reproduced consistently across the other backends, where \textit{Full} reaches the lower-energy basin, and should be interpreted as a backend-specific edge case of over-constrained exploration rather than a case-level failure of the \textit{Full} workflow.

These edge cases illustrate the exploration penalty in the reliability--exploration trade-off. Feedback constraints that improve average reliability can occasionally suppress alternative placements that would have led to lower-energy adsorption configurations. The benefit of closed-loop feedback is therefore conditional. It is largest when the initial adsorption hypothesis is unreliable, smaller when the first proposal already reaches a favorable basin, and occasionally negative when early feedback over-constrains subsequent exploration.

\begin{table}[t]
\centering
\caption{\textit{Full}, \textit{w/o Term}, and \textit{1-Shot} performance on OCD-GMAE62 Tier~1 (62 cases $\times$ 4 backends). $\Delta E = E_{\mathrm{variant}} - E_{\mathrm{Full}}$; positive values indicate higher (worse) energy relative to \textit{Full}. Success rates are calculated over all 248 attempted runs. Mean and median $\Delta E$ are computed over successful same-backend paired comparisons for which both the variant and \textit{Full} have valid molecular adsorption energies. Backend range denotes the maximum--minimum spread across successful LLM backends for each case and is reported only for cases with at least two successful LLM backends.}
\label{tab:ocd62-tier1-full}
\begin{tabular}{lcccccc}
\toprule
Variant & Successful & Success rate & \makecell{Mean \\ $\Delta E$ (eV)} & \makecell{Median \\ $\Delta E$ (eV)} & \makecell{Mean \\ backend range (eV)} & \makecell{Cases with \\ range $\le$ 0.01\,eV}\\
\midrule
\textit{Full}      & 245 & 98.8\% & $+0.000$ & 0.000 & 0.183 & 24/61 \\
\textit{w/o Term} & 247 & 99.6\% & $-0.039$ & $0.000$ & 0.213 & 24/62 \\
\textit{1-Shot}    & 222 & 89.5\% & $+0.329$ & 0.067 & 0.316 & 23/58 \\
\bottomrule
\end{tabular}
\end{table}

Table~\ref{tab:ocd62-tier1-full} supports the pattern observed on AA20: the \textit{1-Shot} penalty is substantial and recurrent, whereas the \textit{w/o Term} variant remains close to \textit{Full} while permitting modest additional exploration. The \textit{Full} configuration also compresses the mean LLM-backend energy range from 0.316~eV under \textit{1-Shot} to 0.183~eV, corresponding to a 1.73-fold reduction among cases with at least two successful LLM backends. This improvement indicates stronger cross-backend agreement under closed-loop feedback, although agreement remains case-dependent rather than universal. The heuristic baseline on OCD-GMAE62 further supports the breadth--reliability trade-off observed on AA20: it identifies more favorable raw MLFF adsorption energies in many cases, but requires a much larger relaxation budget, with a median of 61.5 relaxations per case and a mean of 66.03 (range 19--126), compared with the \textit{AdsMind Full} average of 4.67 relaxations per case.

\subsubsection{Tier~2: Stability audit}

Hosted LLM services are black-box stochastic systems whose outputs may vary across repeated runs, even under fixed prompts and identical runtime settings. Tier~2 quantifies this run-to-run variability by repeating 12 OCD-GMAE62 cases with $N=3$ selected repeated runs under the fixed selection protocol described in the Supporting Information. The dominant pattern for the \textit{Full} configuration is high repeat stability: 72.9\% of backend--case groups (35/48) return selected adsorption energies with a run-to-run range below 0.001~eV. This result indicates that feedback-conditioned replanning, \textit{FORBID} constraints, and convergence-based termination reduce the propagation of LLM variability into the final selected molecular adsorption energies. The residual stochastic tail contains non-identical backend--case groups with energy ranges that can reach 0.5--1.0~eV, suggesting that stochastic proposal differences can occasionally direct the workflow toward distinct local adsorption basins. These residual differences mean that single-run energies should be interpreted as representative workflow outcomes under a fixed protocol rather than unique global solutions.

Run-to-run stability depends strongly on the feedback architecture, although the ordering among ablations is not captured by a simple median-range ranking because many backend--case groups reproduce exactly. Under the fixed $N=3$ repeat protocol, \textit{Full} has the smallest mean run-to-run adsorption-energy range among the main variants (0.033~eV over 48 backend--case groups) and the highest exact-repeat fraction (35/48 below 0.001~eV; 38/48 within 0.01~eV). Removing convergence-based termination increases the mean range to 0.108~eV, while the open-loop \textit{1-Shot} baseline has fewer valid backend--case groups (40/48) and a mean range above \textit{Full} among those valid groups (0.089~eV). Thus, closed-loop feedback improves repeat stability both by narrowing the energy spread of valid repeats and by reducing failure-driven missing outcomes. The two additional variants, \textit{w/o Slip} and \textit{w/o Forbid}, show variant-dependent stability behavior, consistent with partial overlap among the closed-loop mechanisms (Supporting Information).

Taken together, the OCD-GMAE62 results clarify when closed-loop feedback is most beneficial. Across the full Tier~1 benchmark, iterative feedback improves workflow reliability (98.8\% versus 89.5\% success rate for \textit{Full} and \textit{1-Shot}, respectively) and lowers the selected molecular adsorption energy relative to the open-loop baseline, with a mean \textit{1-Shot} penalty of $+0.329$~eV relative to \textit{Full}. Case-level analysis shows that these benefits are largest in chemically complex settings, including multi-element surfaces and multi-atom adsorbates, where first-pass adsorption hypotheses are more likely to be unreliable. In easier cases where the first proposal already reaches a favorable adsorption basin, \textit{1-Shot} can remain close to \textit{Full}. Conversely, in rare edge cases, feedback constraints can over-constrain subsequent exploration after an initially favorable proposal. Thus, closed-loop feedback provides the greatest value when first-pass adsorption hypotheses are unreliable, while its guardrails introduce a measurable exploration trade-off.

\section{Discussion}

\subsection{What Closed-Loop Reasoning Contributes}

The central finding of this study is that closing the feedback loop between LLM-based adsorption planning and MLFF-based structural relaxation improves adsorption-configuration search along three complementary dimensions: reliability, feedback-conditioned error correction, and interpretability. In practice, these benefits also translate into improved relaxation-budget control relative to broad enumeration. Prior work has used active learning as a paradigm for adaptive, feedback-guided sampling~\cite{settles2012active} and MLFFs to reduce per-configuration evaluation expense~\cite{lan2023adsorbml}. AdsMind builds on these ideas by combining MLFF-accelerated relaxation with closed-loop diagnostic feedback in a single adsorption-search framework. In this framework, relaxed structures are not only final outputs but also feedback signals for subsequent planning.

\textbf{Reliability.} Across both benchmarks and all four LLM backends, \textit{AdsMind Full} achieves consistently high success rates while counting natural chemical failures, such as dissociation or rearrangement during structural relaxation, in the success-rate denominator. This accounting makes the reported reliability closer to a practical workflow-level measure than a filtered success rate over only chemically clean cases. In the six-system DFT validation subset, AdsMind preserves the qualitative adsorption-energy sign in all tested cases and shows closer agreement with DFT than the reported open-loop Adsorb-Agent outputs. Separately, relative to the \textit{1-Shot} ablation, AdsMind reduces cross-backend energy dispersion on both benchmarks, indicating improved agreement among heterogeneous hosted LLM backends for valid molecular adsorption energies. Although this does not eliminate backend dependence, it reduces sensitivity to the specific LLM used and provides a more stable basis for downstream adsorption-energy comparisons. This feature is important for LLM-based architectures that rely on hosted, black-box models that may be subject to unannounced updates.

\textbf{Feedback-conditioned correction.} The Planner conditions each new proposal on the preceding relaxation outcomes, including failed placements, \textit{Chemical Slip} events, dissociation warnings, and previously successful configurations. This history-conditioned replanning converts physical feedback into subsequent action choices, allowing the system to correct chemically or geometrically inconsistent hypotheses rather than merely enumerate static candidates.

\textbf{Interpretability.} The closed-loop architecture yields physically grounded diagnostics, such as \textit{Chemical Slip} events, \textit{FORBID} constraints, and planned-versus-relaxed site mismatches that make the search process inspectable rather than just reducing it to a final adsorption energy. Although such diagnostics can in principle be computed post hoc, AdsMind uses them as internal feedback signals during planning, which distinguishes it from open-loop search protocols. By stratifying \textit{Chemical Slip} rates across surface families, we obtain a quantitative diagnostic proxy for where LLM-generated adsorption hypotheses are geometrically fragile, complementing text-based chemistry benchmarks such as ChemBench~\cite{mirza2024chembench}. \textit{Chemical Slip} tracking therefore reveals whether an LLM-proposed binding mode remains consistent after structural relaxation, which is an important capability for autonomous scientific discovery workflows in computational catalysis.

\textbf{Auditability.} When a hosted LLM proposes the calculations, a user may reasonably ask whether the conclusions of the workflow can be checked or must be taken on trust. In AdsMind, the LLM proposes symbolic hypotheses but never sets an energy. Every reported energy comes from an MLFF relaxation of an explicit structure, which can be re-examined with another force field or with DFT, as in our validation of six AA20 systems. For every attempt, each run stores the plan together with the Planner's stated reasoning, the relaxed structure, and the Analyzer's diagnostics, and the Summarizer report tabulates planned against relaxed sites, so a user can trace why the search ended where it did. The attempt-level records of all benchmark runs reported here, with every plan, the Planner's reasoning, and the Analyzer's diagnostics, are available in the project repository. We therefore regard AdsMind as a tool for proposing and triaging adsorption configurations: its selected structures are well-defined starting points whose energies should be confirmed with first-principles calculations before they support quantitative conclusions.

\subsection{Limitations}

\textbf{Reliability--exploration trade-off.} AdsMind's closed-loop design prioritizes reliable convergence and relaxation-budget control over exhaustive exploration. Because the \textit{Full} workflow typically terminates after a small number of MLFF relaxations and uses the \textit{FORBID} directive to discourage surface-element combinations that have already slipped, it can occasionally converge to a local basin on the potential-energy surface (PES) rather than the lowest-energy basin accessible within the MLFF-defined search space. This limitation is most visible for structurally complex adsorbates, such as CH$_2$CH$_2$OH and OCHCH$_3$, where the selected adsorption energy can remain up to $\sim$1~eV higher than that identified by broad-sampling baselines. This gap suggests that adaptive multi-seed exploration could be useful in difficult cases, but the current workflow does not fully close the residual search-breadth gap. In addition, the \textit{FORBID} directive could over-prune the search space if a surface-element combination that slipped once still admitted a favorable placement. The ablations show little sign of this. Removing the directive leaves the mean selected energy nearly unchanged (mean $\Delta E$ of $-0.024$~eV on AA20 and $+0.020$~eV on OCD-GMAE62, with 63/80 and 182/244 runs within $\pm0.05$~eV of \textit{Full}; Tables~S9 and~S11), and removing \textit{Chemical Slip} feedback together with it gives similar results ($-0.021$ and $+0.021$~eV; 59/80 and 171/245 runs). Individual runs shift in both directions: with a 0.01~eV threshold, 18 of the 80 AA20 runs end lower without \textit{FORBID} and 10 end higher, and on OCD-GMAE62, 40 of the 244 runs end lower and 50 higher. In none of the 30 paired runs in which the variant without \textit{FORBID} ended more than 0.1~eV lower did its best configuration come from re-planning a surface-element combination that had already slipped, so these differences reflect the stochasticity of the search rather than combinations excluded by the directive. For GPT-5.4 on OCD-GMAE62 case~008, where \textit{1-Shot} returns a structure 0.99~eV lower than \textit{Full}, the \textit{Full} and \textit{w/o Forbid} runs make the same three plans, so this gap does not come from the directive either. These behaviors reflect the same architectural trade-off: closed-loop guardrails improve average reliability while reducing worst-case search breadth. Future work should therefore investigate adaptive termination criteria and context-sensitive constraints that can be relaxed when the workflow has insufficient evidence for convergence.

A practical benefit of this trade-off is a reduced MLFF relaxation budget. \textit{AdsMind Full} uses 4.11 MLFF relaxations per case on AA20 and 4.67 on OCD-GMAE62 on average, compared with mean heuristic-enumeration budgets of 56.85 and 66.03 attempted MLFF relaxations per case, respectively, under the same MACE-MP-0 (small) structural relaxation protocol. This relaxation-budget comparison refers specifically to physical-backend calls rather than LLM usage. This advantage should be interpreted as a controlled-search benefit rather than evidence of exhaustive global-minimum discovery.

\textbf{Hypothesis schema.} The Planner states each hypothesis in a small, fixed vocabulary: one of three canonical site types, the element symbols of the intended surface binding atoms, and one or two adsorbate binding atoms. This schema makes every hypothesis checkable before simulation and directly comparable with the relaxed outcome, but it cannot express every adsorption motif. It does not distinguish subsites such as fcc and hcp hollows or name specific surface atoms, and it leaves the placement and orientation of the adsorbate to the \textbf{Executor}, which samples up to eight sites and four conformers per site by default. Binding through three or more adsorbate atoms, co-adsorption and coverage effects, and adsorbate-induced surface reconstruction lie outside the schema. Reconstruction can still emerge during relaxation, because the upper slab layers relax freely, but the Planner cannot propose it. On such surfaces, the closed-loop mechanisms could also mislead the search. Surface atoms that move during relaxation change the set of atoms bonded to the adsorbate. A \textit{Chemical Slip} may then be flagged, and \textit{FORBID} may discourage a surface-element combination that is in fact favorable. We tested this risk on two surfaces with experimentally documented adsorbate-induced reconstruction, methylthiolate on Au(111)~\cite{maksymovych2006adatom} and atomic carbon on Ni(100)~\cite{klink1993clock}. For three backends with five repeats each, a search with all three mechanisms disabled but the five-attempt budget kept was compared with \textit{1-Shot} and \textit{Full} (Supporting Information, Table~S15). \textit{Chemical Slip} flagged 27 of 42 \textit{Full} attempts on Au(111) and 41 of 42 on Ni(100), where carbon in the fourfold hollow also bonds to the nickel atom below. All 90 runs still ended within 0.05~eV of the lowest adsorption energy found on their surface, and the best \textit{Full} configuration came from a slipped attempt in 8 of 15 runs on Au(111) and in all 15 on Ni(100). The flags therefore did not discard favorable configurations, and \textit{TERMINATE} let \textit{Full} stop after fewer attempts. These runs hold one adsorbate per $3\times3$ surface cell, so they probe the local displacements induced by adsorption, up to 0.33~\AA\ for gold atoms bonded to sulfur, rather than the extended reconstructed phases that form at higher coverage. Part of the hypothesis could instead be learned from short MLFF calculations. The \textbf{Executor} already ranks candidate placements within each hypothesis by single-point energies, across matching sites that differ in local geometry and across several orientations of the adsorbate. A warm-up stage could extend this screening to the bonding environments themselves, including positions away from the enumerated sites and symmetry-breaking displacements of the adsorbate, for example with the short Langevin warm-up already available in the released code. It would pass the most favorable site types and surface-element combinations to the Planner as candidates, so that the LLM chooses among physically ranked options rather than proposing them from scratch, and the search could also reach minima that lie away from the enumerated sites. Evaluating this hybrid design requires new benchmark runs and is left to future work.

\textbf{Force-field and chemical-domain scope.} The primary evaluation uses MACE-MP-0 (small) as the MLFF backend and is limited to the crystalline slab surfaces and single-adsorbate settings represented in AA20 and OCD-GMAE62. Sensitivity checks that change the MLFF configurations to MACE-MP-0 (large) with CUDA execution, float64 precision, and dispersion enabled produce a mean absolute energy shift of 1.255~eV on AA20, indicating that absolute adsorption energies, and in some cases, relative rankings can depend on the chosen MLFF protocol. Although the six-system DFT validation subset shows that AdsMind correctly preserves the qualitative adsorption-energy sign and improves agreement with DFT relative to the reported open-loop Adsorb-Agent outputs, its quantitative accuracy remains inherently bounded by the physical fidelity of the underlying MLFF. Relevant sources of error include the treatment of dispersion, charge transfer, magnetic effects, and strongly correlated bonding.

Extending AdsMind to defective surfaces, nanoparticles, high-coverage regimes, co-adsorption, solvent environments, or electrochemical conditions will therefore require additional validation against appropriate benchmarks, such as QUASAR~\cite{yang2026quasar}, and may also require expanded site definitions, binding-mode schemas, and validation rules. The core feedback principle---using relaxed structures as diagnostic signals for subsequent planning---is backend-agnostic and conceptually transferable, but its chemical reliability must be re-established whenever the force field, surface class, or adsorption regime changes.

\textbf{LLM backends.} All results reported here were obtained with four hosted frontier models, and part of what the benchmarks measure is their quality as Planners. The backend interface of AdsMind also accepts locally served open-weight models through Ollama and Hugging Face, so the workflow can be run end to end without paid API access, but we have not benchmarked such models. Whether smaller local models can sustain schema-valid, index-consistent planning, and how often their hypotheses slip, remain open questions. The diagnostics introduced here are suited to answering them, because \textbf{Validator} rejection rates and \textit{Chemical Slip} rates measure planning errors directly rather than only through the final energy.

\textbf{Run-to-run stochasticity.} The closed-loop architecture improves run-to-run stability under fixed runtime settings, but it does not eliminate stochasticity from hosted LLM-based workflows. In the Tier~2 stability audit ($N=3$ selected repeated runs), 72.9\% of \textit{Full}-configuration backend--case groups (35/48) return selected adsorption energies with a run-to-run range below 0.001~eV. This result indicates that feedback-conditioned replanning, \textit{FORBID} constraints, and convergence-based termination reduce the extent to which LLM stochasticity propagates into the final molecular adsorption energies. Because many backend--case groups reproduce exactly, median ranges are not a robust discriminator among variants. Using mean run-to-run range and valid-group counts, \textit{Full} is the most stable main variant (0.033~eV over 48 backend--case groups), whereas removing convergence-based termination increases the mean range to 0.108~eV and the open-loop \textit{1-Shot} baseline retains fewer valid backend--case groups (40/48) with a higher mean range than \textit{Full} among those valid groups (0.089~eV). These residual variations should not be treated as numerical noise alone. In some cases, stochastic proposal differences drive the workflow toward distinct but physically valid local adsorption basins. Therefore, single-run outputs should be interpreted as representative outcomes of the fixed workflow protocol rather than unique global solutions.

\textbf{Within-session memory and cross-case transfer.} AdsMind maintains within-session memory that guides per-case search through \textit{Chemical Slip} signals, \textit{FORBID} constraints, and prior relaxation outcomes. However, this memory is not persisted across adsorption-search sessions. Each new surface--adsorbate pair begins without persistent information about failure patterns or favorable binding motifs encountered in previous cases. For example, if a specific site family repeatedly slips on Pd-bearing intermetallic surfaces, the current implementation must rediscover this behavior independently in each relevant case. Extending the current within-session memory to a persistent cross-case knowledge store could reduce redundant exploration as the workflow is applied to larger surface--adsorbate libraries. Such an extension would require additional mechanisms for knowledge representation, retrieval, uncertainty tracking, and validation, because transferred heuristics may be system-dependent and may become misleading outside the chemical domain in which they were observed.

\section{Conclusion}

We introduce AdsMind, a closed-loop multi-agent framework that couples LLM-driven adsorption-hypothesis generation with iterative MLFF relaxation feedback. This design addresses a key limitation of open-loop LLM agents: the inability to use structural relaxation outcomes to detect, diagnose, and revise chemically or geometrically inconsistent adsorption hypotheses across iterations. Evaluated across 20 AA20 cases and 62 OCD-GMAE62 cases using four different LLM backends, AdsMind simultaneously delivers high search reliability, feedback-conditioned self-correction, and interpretable diagnostic artifacts. These properties are difficult to obtain from prior open-loop adsorption-search methods. By deliberately occupying the reliability-first, low-relaxation-budget regime, AdsMind serves as a practical complement to broader enumeration strategies that prioritize exhaustive search breadth.

Future work should integrate AdsMind's closed-loop error-recovery mechanisms with broader initial enumeration or adaptive multi-seed exploration, while extending validation to more complex adsorption environments such as co-adsorption, solvation, electrochemical conditions, and defective surfaces. More broadly, the \textit{Chemical Slip} diagnostic provides a general strategy for connecting LLM-generated symbolic adsorption hypotheses with physically relaxed structures. By quantifying the mismatch between LLM-planned and MLFF-realized binding modes, this diagnostic offers a practical route toward more interpretable and physically grounded LLM-assisted autonomous scientific discovery workflows for computational catalysis.

\section*{Data Availability}

All benchmark inputs, frozen run configurations, execution scripts, and curated summary outputs generated in this study are publicly available in the AdsMind repository at \url{https://github.com/AI4QC/AdsMind}. 
The GitHub repository provides the AA20 and OCD-GMAE62 benchmark manifests, frozen model and runtime configurations, scripts for reproducing the benchmark runs and summary analyses, full summary tables for four LLM backends across the primary five variants, additional AA20 per-agent ablations, and performance comparisons against the Heuristic and Adsorb-Agent baselines. The repository also includes the $N=3$ repeated-run stability audit data for OCD-GMAE62 and, for every benchmark run, the attempt-level records (plans with the Planner's reasoning, Analyzer diagnostics, and history entries), so that each search trajectory can be inspected without querying an LLM.

Since hosted LLM services are subject to unannounced model and infrastructure updates, the frozen run configurations, including model identifiers, MACE-MP-0 settings, and runtime switches, are intended to support reproducibility of the reported summaries but may not guarantee exact numerical parity in future reruns. Rerunning the hosted-model backends requires API keys from the respective providers, whereas the locally served backends (Ollama and Hugging Face) run without any API key. MACE-MP-0 checkpoints are publicly available~\cite{batatia2023foundation}.

\section*{Acknowledgments}

This work was financially supported by the Swiss National Science Foundation (SNSF) through the National Centre of Competence in Research (NCCR) Catalysis (grant no. 225147). The authors thank the EPFL Scientific IT and Application Support team for computational support. We thank HKUST Fok Ying Tung Research Institute and National Supercomputing Center in Guangzhou Nansha Sub-center for computational resources. Z.Z. acknowledges personal support from the Hong Kong Special Administrative Region Government Scholarship Fund and HKUST Study Abroad Funding Support. Z.Z. also thanks the Laboratory of Artificial Chemical Intelligence (LIAC) at École Polytechnique Fédérale de Lausanne (EPFL) for hosting him during his exchange semester. The authors thank Janghoon Ock for helpful correspondence on reproducing the 20-case Adsorb-Agent AA20 surface-generation protocol used in this study.

The authors used generative AI and AI-assisted tools, including large language models such as ChatGPT, Claude, Gemini, and DeepSeek, as well as Codex, Claude Code, Grammarly, for grammar polishing, language editing, and minor phrasing improvements during the preparation of the manuscript. This use was limited to manuscript preparation and was separate from the LLM-agent methodology described in the Methods section. All scientific analyses, interpretations, and conclusions were solely produced by the authors. The authors reviewed and edited all AI-assisted text and take full responsibility for the content of the manuscript.

\section*{Conflict of Interest Disclosure}

The authors declare no competing financial interests.

\bibliography{journal_abbrev,refs}

\end{document}